\documentclass{aa}
\usepackage{epsfig}

\begin{document}
\thesaurus{ }
\title{A photometric survey for variable stars in the globular cluster M5}  
\author{ ~J. Kaluzny\inst{1,2}, ~I. Thompson\inst{3},
~W. Krzeminski\inst{1,3}, ~W. Pych\inst{2} }
\offprints{J. Kaluzny, e-mail: jka@sirius.astrouw.edu.pl}
\institute{
Copernicus Astronomical Center, Bartycka 18, 00-716 Warsaw, Poland
\and
Warsaw University Observatory, Al. Ujazdowskie 4, 00-478
Warsaw, Poland
\and
Carnegie Institution of Washington, 813 Santa Barbara Street, Pasadena,
CA 91101, USA}

\date{Received 1 April, 1999, Accepted 20 July, 1999}

\maketitle
\markboth{J. Kaluzny  et al. Variables in M5}{}
\begin{abstract}
We present the results of a photometric survey for variable stars in
the  nearby globular cluster M5. A $14.8\times 22.8$ arcmin field
centered on the cluster was monitored for a total of 37 hours with a
CCD camera mounted on the 1.0-m Swope telescope.  Five new variables
were identified:  four blue stragglers which are SX Phe pulsating
variables, and an eclipsing binary with an orbital period of 0.466 d.
The eclipsing binary lies near the main-sequence turnoff on the cluster
color-magnitude diagram.  We do not confirm the variability of any of
the 10 eclipsing binaries identified in the cluster field by Reid
(1996) and Yan \& Reid (1996).  The dwarf nova  M5-V101 exhibited two
outbursts with full amplitudes of about 2.7 mag during our
observations.  A $V/B-V$ color-magnitude diagram of the surveyed field
was obtained, and a possible extreme HB star located 2 mag below cutoff of
the blue HB was identified.
  
\end{abstract}

\section{Introduction}

NGC~5904=M5={\bf C1516+022} is a  rich globular cluster whose proximity and low 
reddening ($r \approx 7.0$~kpc, $E(B-V)=0.03$; Harris 1996) make it 
an excellent target for detailed studies. We selected this cluster
as one of the  targets for an ongoing survey for eclipsing binaries in
globular clusters (Kaluzny et al. 1997; 
Thompson et al. 1999).
The main goal of this project is to use detached eclipsing binaries
for the determination of ages and distances of globular clusters 
(Paczy\'nski 1997).\\ 

M5 is known to contain more than 120 RR Lyr variables (Clement 1997
priv. comm.).
CCD $VI$ photometry for 49 of them was presented recently by Reid (1996)
who gives  references to earlier papers dealing with RR Lyr
stars in M5. As a side result of our survey we obtained $V$-band light
curves for 65 RR~Lyr stars from M5. These data will be presented in a
separate contribution (Kaluzny et al. 1999). 

Reid (1996) reported  the discovery of 4 probable eclipsing
binaries among the cluster HB/RGB stars. In addition, Yan \& Reid (1996) 
identified 6 short-period  eclipsing binaries located at the
cluster turn-off. Our data  indicate that all  of the eclipsing
variables reported in Reid (1996) and Yan \& Reid (1996) were 
spurious detections. 
  
Drissen \& Shara (1998) used HST images to look for variable stars
in the core of M5. They managed to identify one variable blue straggler, 
a possible eclipsing binary with a period $P\geq 0.47$ d.
Several new RR Lyr stars were also identified in this study. 

M5 is one of only two galactic globular clusters known to harbor
dwarf novae. Variable M5-V101 was first identified as a dwarf nova 
by Oosterhoff (1941). A recent summary of observational studies 
of that variable is given in Hakala et al. (1997). 
During our monitoring of M5 the variable exhibited two outbursts with 
full amplitudes exceeding 2.7 mag in the V band.

\section{Observations and data reduction} 

The time-series photometry of M5 on which this paper is based was
obtained with the 1.0-m Swope telescope at Las Campanas Observatory.
The data were collected during the interval 1997 May 9 -- 1997 Aug 13.
A condensed  log of the observations is given in Table 1.

\setcounter{table}{0}
\begin{table}
\caption[ ]{Observing log }
\begin{flushleft}
\begin{tabular}{cll}
\hline
Date & $N_{V}$ & $\Delta t$  \\
UT   &         &  hours  \\
\hline
May 9 & 31 & 5.5  \\
May 10& 34 & 5.8 \\
May 13& 37 & 5.6 \\
May 14& 37 & 6.5 \\
May 15& 44 & 5.8 \\
May 19& 24 & 2.0 \\
Jun 7 &  9 & 2.0 \\
Jun 8 & 12 & 2.0 \\
Jun 17& 10 & 1.5 \\
Aug 3 & 1  &  - \\
Aug 4 & 1  &  - \\
Aug 13& 2  &  0.2 \\
\hline
\hline
\end{tabular}
\end{flushleft}
\end{table}

A $2048\times 3150$ pixel  SITe CCD  camera was used as the detector,
with a field of view of $14.8\times 22.8$ arcmin at a scale of 0.435
arcsec/pixel.  Due to some technical problems the field of view of the
camera was limited to $14.8\times 14.8$ arcmin during the period from
1997 May 9 through 1997 May 19.  During that run we monitored two
overlapping subfields covering the northern and southern sections of
the cluster. The total area of the monitored field was equal to 314
square arcmin.  Starting in 1997 Jun 7 we monitored just one $14.8\times
22.8$ arcmin field centered on the cluster core. Monitoring was
conducted in the $V$ band with exposure times ranging from 300 to 500
seconds, depending on the seeing. The median value of  the seeing for
all of the images used for this survey was 1.52 arcsec.  Only two
images in the $B$-band were  taken during the 1997 season.  On two
nights in May/June 1998 we collected frames of the cluster field
through the $V$ and $B$ filters using the same instrumentation as in 1997.
These exposures were used to derive the  color-magnitude
diagram for the field monitored in 1997 (see Sect. 4).

Instrumental photometry was extracted with DoPHOT (Schechter et al.
1993) used  in the fixed-position mode. The stellar positions were
derived  from the  reduction of "template" images, the best individual
images obtained for a given sub-field.  A total   of 15839 stars with
$V<20.8$  were checked for variability.  The quality of the derived
photometry is illustrated in Fig. 1 in which we plot the rms
deviation versus the average magnitude for stars from the northern
sub-field.  This plot includes 14847 stars with $13.5<V<20.8$ whose
light curves contain at least 75 points.  Images of  stars with
$V<13.5$ were generally over-exposed, and the resulting photometry 
was poor.  To
select potential variables we employed three methods as described in
some detail in Kaluzny et al. (1996).  Light curves showing possible
periodic signals or smooth changes on time scales of  days to weeks
were selected for further examination. A total of 71 certain variables
were identified: 65 RR Lyr stars, all of them already known, four
previously unknown SX Phe stars,  one  previously unknown eclipsing
binary, and a previously known cataclysmic variable (M5-V101).  In
Table 2 we list the equatorial coordinates for the 5 newly identified
variables.  Finding charts  are shown in Fig. 2.

\setcounter{table}{1}
\begin{table}
\caption[ ]{Equatorial coordinates for 5 newly identified
variables}
\begin{flushleft}
\begin{tabular}{lll}
\hline
ID & RA(2000) & Dec(2000)  \\
   & h:m:sec & deg:$\arcmin$:$\arcsec$  \\
\hline
M5-NV1 & 15:18:22.76  & 02:02:49.3 \\
M5-NV2 & 15:18:11.40  & 02:00:40.6 \\
M5-NV3 & 15:18:40.04  & 01:58:41.6 \\
M5-NV4 & 15:18:41.02  & 01:56:29.6 \\
M5-NV5 & 15:18:39.10  & 02:11:41.0 \\
\hline
\hline
\end{tabular}
\end{flushleft}
\end{table}

\subsection{Transformation to the standard system}

Observations of Landolt (1992) standards collected on
several nights during the 1997 observing season were used to
establish transformations from the instrumental to the 
standard BV system. The following relations were obtained:
\begin{eqnarray}
v = a_{1} + V -0.010\times (B-V) + k_{v}\times X\\
b-v = a_{2} + 0.975\times (B-V) + k_{bv}\times X
\end{eqnarray}
For the 1998 season we derived:
\begin{eqnarray}
v = a_{1} + V -0.006\times (B-V) + 0.126\times X\\
b-v = a_{2} + 0.953\times (B-V) + 0.139\times X
\end{eqnarray}
The extinction coefficients in Eqs. 3 and 4 correspond to the night 
of May 29/30, 1998,
on which we obtained data used to construct the color-magnitude diagram 
for the field monitored for variables (see Sect. 4).
While transforming the 1997 light curves of variables from the instrumental 
v magnitudes to the standard V magnitudes we used a simplified relation:
\begin{eqnarray}
V = v + 2.435
\end{eqnarray}
This approach was required since no observations of the light curves in the
B band were obtained. 
The constant offset was derived using photometry of non-variable 
stars with $14.5<V<17.0$ and $0.1<B-V<0.75$. 
The adopted procedure leads to systematic errors not exceeding
0.01 mag.\footnote{The color term  in Eq. 1  equals  0.010.
All variables discussed in this paper 
have average colors in the range $0.20<B-V<0.43$.
For RR Lyr variables changes of the $B-V$ color 
seldom exceed 0.3 mag over the whole pulsational cycle.}

\subsection{Comparison with previous photometry}
The field covered by our survey overlaps the region studied by
Sandquist et al. (1996). Unfortunately, it does not include  the field
observed by Stetson and Harris (1988) and more recently by Johnson \&
Bolte (1998).  Comparison of our BV data with Sandquist et al. (1996)
is shown in Fig. 3 (differences are given as "ours" - "theirs").  The
mean differences for stars with $V<17.5 $ are $0.005\pm 0.015$ and
$-0.021\pm 0.017$, for $V$ and $B-V$, respectively.  A systematic trend
of $\Delta V$ residuals with increasing $V$ magnitude can be noted in
Fig. 3. A similar trend was noted by Johnson \& Bolte (1998) who
compared their $VI$ data with Sandquist et al. (1996).

\section{Results for variables}

Our survey resulted in the detection of five new variables: four SX Phe
type stars and an eclipsing binary.  Figure 4 shows the location of
these objects  on the cluster color-magnitude diagram (CMD). For the
SX~Phe stars the  marked position corresponds to the intensity averaged
magnitude and to the color measured near maximum light. For the
eclipsing binary  the marked position corresponds to the magnitude and
color measured near the center of the secondary minimum at the orbital
phase 0.52.  The position of the dwarf nova M5-V101 is also marked in
Fig. 4, and corresponds to the magnitude and color measured for that
highly variable object on HJD=2450963.57 (May 30, 1998 UT).
   
The phased light curves of the four SX Phe variables  are shown 
in Fig. 5 and the light curve of the eclipsing binary is shown in
Fig. 6. Some basic photometric parameters of the newly identified
variables are listed in Table 3. The periods of variability were 
derived using an
algorithm based on the {\it analysis of variance} statistic
(Schwarzenberg-Czerny 1989).

\setcounter{table}{2}
\begin{table}
\caption[ ]{Light curve parameters for newly identified variables. The
last column gives intensity averaged magnitude for SX Phe stars.} 
\begin{flushleft}
\begin{tabular}{clcccc}
\hline
ID & $P[{\rm day}]$ & $B-V$ & $V_{max}$ & $V_{min}$ & $A_{V}$ \\
\hline
M5-NV1 & 0.04213  & 0.20    & 17.42 & 17.53 & 17.48 \\
M5-NV2 & 0.04130  & 0.28    & 16.75 & 16.77 & 16.76 \\
M5-NV3 & 0.04166  & 0.19    & 16.99 & 17.09 & 17.04 \\
M5-NV4 & 0.04745  & 0.24    & 16.89 & 16.95 & 16.92 \\
M5-NV5 & 0.4663   & 0.43    & 18.91 & 20.07 &       \\
\hline
\hline
\end{tabular}
\end{flushleft}
\end{table}

\subsection{SX Phe stars} 

The SX Phe variables are located in the blue straggler region of   the
cluster CMD.  The observed properties of these variables allow  their
classification as SX Phe stars.  Fig. 7 shows the positions of the
variables in a period vs. absolute magnitude diagram. The
standard relations for SX Phe stars pulsating in the fundamental mode
and in the first overtone are also shown (McNamara 1997, see also
McNamara 1995). The plotted relation for stars pulsating in the
fundamental mode is 
\begin{eqnarray}
M_{V} = -3.725 \times{\rm  Log} P_{0} - 1.933
\end{eqnarray}
and the same
relation is used for stars pulsating in the first overtone assuming
$P_1/P_0$ = 0.778.
We also plot the $P-L-[Fe/H]$ relations for fundamental and first
overtone pulsators from Nemec et al. (1994), assuming metallicity of
$[Fe/H]=-1.4$.
The absolute magnitudes for M5 variables were
calculated assuming a distance modulus of $(m-M)_{V} = 14.33$ (Harris
1996).  It is clear that the only variable in Fig. 7 close to the
standard relation for the fundamental mode pulsators is M5-NV1.  It is
likely that the 3 remaining SX Phe stars are first overtone pulsators.
The small amplitudes and the symmetry of their light curves support
such a hypothesis, though as McNamara (1997) points out, classification of
a star as a first overtone pulsator based only on these criteria is
questionable.  However, even assuming that the  3 brightest of the
4 identified SX Phe stars are first overtone pulsators, we still face
the problem of an apparent over-luminosity of all 4 variables 
most especially for the Nemec et al. (1994) relations.   
One possible solution to this problem is adoption of a lower distance
modulus for the cluster, $(m-M)_{V}\approx 14.1$, a value entirely
discrepant with that adopted by Harris (1996) for this well studied
cluster.  This would also imply a very low luminosity for the cluster
RRab  variables of $M_{V}\approx +1.00$ (the average $V$ magnitude for
RRab stars in M5 is 15.062 (Kaluzny et al. 1999)).  Another possibility
is that discovered SX Phe stars are not members of M5,
{it although
M5 is at a galactic latitude of $b=46.8$ deg, and this expalnatgion
implies un unreasonably high field surface density for SX Phe stars}.
One way to
resolve  this puzzle would be to measure the radial velocities for
these variables to check their membership status. A more likely 
explanation is that the absolute magnitude - period relations are
poorly known for stars of such short periods. Indeed, McNamara's
calibration based on Hipparcos measurements has only five stars with
periods below log $P = -1.1$, and the scatter in this region of the
calibration is significant. It is hoped that our survey for variable
stars in other globular clusters will lead to the discovery of significant
populations of SX Phe stars, leading to a more secure calibration.

\subsection{Eclipsing binary stars}

Variable M5-NV5 is potentially an important object as it may prove to
be a detached  eclipsing binary composed of two upper-main sequence
stars from M5. As such this binary could be used for measuring the
distance to the cluster. In addition, the determination of the masses,
radii and luminosities of the components of this system is of great
interest.  However, the  present data  are not sufficient for a precise
determination of parameters of M5-NV5.

To get some hints about the membership status and configuration of
M5-NV5, we analyzed the light curve presented in Fig. 6.  The light
curve was solved using the MINGA package (Plewa 1988) which uses the
Wilson-Devinney (W-D) code (Wilson \& Devinney 1971, Wilson \& Sofia
1976 ) to generate synthetic light curves.  The code was used in a mode
in which the basic parameters determining the shape of  the synthesized
light curve are the mass ratio $q = M_{2}/M_{1}$, the potentials
$\Omega_{1}$ and $\Omega_{2}$ (for a given $q$ these determine the
relative radii $r_{1}$ and $r_{2}$), the relative luminosities $L_{1V}$
and $L_{2V}$, and the inclination of the orbit $i$.  During the
analysis we fixed the temperature of the primary component at
$T_{1}=6300$~K. According to the derived solution the secondary eclipse
is total or very close to total. Hence at an orbital phase of 0.52, at
which we measured $B-V$, only the primary component is visible.
Adopting $E(B-V)=0.03$ we obtain $(B-V)_{1}=0.40$ which in turn implies
$T_{1}=6300$ for an adopted $[{\rm Fe/H}]=-1.1$ (Alonso et al.
1996; Harris 1996).  As we have no information about the
mass ratio of the system we calculated  a grid of light
curve solutions for a set of values of $q$. The results are summarized
in Table 4.

\setcounter{table}{3}
\begin{table}
\caption[ ]{Light curve solutions for M5-NV5.}
\begin{flushleft}
\begin{tabular}{ccccc}
\hline
$q$ &  $i$  & $L_{1V}/L_{2V}$ & $r_{1}$ & $r{_2}$\\
\hline
0.50 &89.2 &3.68 & 0.32& 0.27\\
0.60 &88.9 &3.78 & 0.32& 0.27\\
0.70 &87.2 &3.82 & 0.31& 0.27\\
0.80 &87.8 &4.07 & 0.32& 0.27\\
0.90 &85.9 &3.45 & 0.30& 0.28 \\
\hline
\hline
\end{tabular}
\end{flushleft}
\end{table}
All of the solutions in Table 4 imply that the system is detached with
the secondary eclipse being total or very close to total.  Based on a
$\chi^{2}$ criterion, the best solution was obtained for $q=0.70$, but
solutions derived for other values of $q$ are only marginally worse.
This is not surprising considering the limited accuracy of the
photometry and the detached configuration of the binary.  To illustrate
the quality of the solutions we plot in Fig.  8 the observed light
curve with the synthetic light curve corresponding to $q=0.70$.
Despite our inability to derive the mass ratio of the system it is safe
to state that M5-NV5 is a detached binary with an inclination $i>86$
deg and  that the luminosity ratio $L_{1V}/L_{2V} > 3.4$.  The position
of the primary component of M5-NV5 on the cluster CMD (see Fig. 4) is
consistent with the hypothesis that it is a main-sequence star
belonging to the cluster. Finally, we note that the derived values of
$L_{1V}/L_{2V}$  and  $r_{1}/r{_2}$ favor solutions with $0.7\leq q
\leq 0.8$ (if we assume that both components are main sequence stars).
If the system  contains two cluster main sequence stars with
masses $m_{1}\approx 0.7 M_\odot$ and $m_{2} \approx 0.5 M_\odot$ then
the expected semi-amplitudes of their radial velocity curves are
$K_{1}=121$ and $K_{2}=170$ km/sec. In this case the velocity curve of
the binary would be relatively easily measured with a medium resolution
spectrograph on a large telescope.

In a paper devoted mainly to a photometric analysis of RR Lyr stars in
M5, Reid (1996) reported the identification of 4 eclipsing binaries --
2 located on the horizontal branch  and 2 located on the giant branch.
Subsequently Yan \& Reid (1996) used the same observational material
to identify 6 short-period eclipsing binaries located on or near the
main-sequence of the cluster.  None of these 10 candidate variable
stars  was selected from our data base during our  search for variable
objects. Our light curves for these stars do not show any sign of
variability of the type reported by Reid (1996) and Yan \& Reid
(1996). At our request Dr. Reid checked his CCD frames of M5. It turned
out that the variability of the 4 candidate eclipsing binaries reported
by Reid (1996) was due to occasional passages of the stellar images
through defective columns of the detector. Our light curves for the 6
candidate variables reported by Yan \& Reid (1996) show scatter
ranging from 0.04 mag for V6 and V5 to 0.08 mag for V2.  None of these
light curves show any modulation with the periods or at the amplitudes
reported by Yan \& Reid (1996), nor with any other periods shorter than the
time-base of our observations.  This is demonstrated in Fig. 9 in which
we show our light curves for stars V1, V2 and V3  from Yan \& Reid
(1996) phased with the periods listed in that paper.

\subsection{The cataclysmic variable M5-V101}
M5-V101 is one of two  cataclysmic variables which were known in
globular clusters in the pre-HST era. It was discovered by Oosterhoff
(1941), who suggested that it is a dwarf nova.  That suggestion was
later confirmed by  spectroscopic as well as photometric observations
(Margon et al. 1981; Shara et al. 1987). Detection
of the X-ray counterpart of M5-V101 was reported by Hakala et al.
(1997).

In Fig. 10 we present the $V$-band light curve of M5-V101 based on our
observations. The variable displayed two outbursts with full amplitudes
of about 2.7 mag during our observations in 1997.  On the night of May 30
1998 we observed M5-V101 at $V=20.27\pm 0.04$ and $B-V=0.10 \pm 0.08$
(HJD = 245 0963.57).  The occurrence of at least two outbursts on a
time scale of 100 days in 1997 as well as the detection by Shara et
al. (1987) of a single outburst during a two-week long monitoring period
indicates a relatively short duty cycle for M5-V101.  Examination of
the nightly light curves from  the 1997 season reveals occasional variability
with a full amplitude of about 0.4 mag on a characteristic time  scale of
about 3.4 hours.  Spectroscopic data presented by Naylor et al. (1989)
suggest that the orbital period of M5-V101 is longer than 3.5 hours.

\section{The color-magnitude diagram}
An extensive and detailed photometric study of M5 based on CCD BVI
data was published  by Sandquist et al. (1996).  Even more recently
Johnson \& Bolte (1998) reported VI data for a field in the outer
part of the cluster.  As a by product of our program we obtained a
medium-deep $V/B-V$ CMD for the $14.8\times 22.8$ arcmin field centered
on the cluster, shown in Fig. 4.  This CMD is based on the following
set of images:  $B$ -- 60~s, 150~s, 260~s; V -- 60~s, $4\times 150$~s.
The observations  were  obtained on the nights of 1998  May 30 and June
3. The seeing for these frames ranged from 1.13 to 1.35 arcsec. 
Photometry was extracted using Daophot/Allstar package (Stetson 1987).
Measurements of stars with relatively poor quality (large sigma for a
given magnitude and/or a large value of Stetson's CHI parameter) were
rejected from the analysis.  Known variables listed in Clement (1997,
priv. comm.)
or in Sandquist et al. (1996) are not plotted in Fig. 4.  The CMD
shows several features discussed  in some detail by Sandquist et al.
(1996). In particular we note the presence of a star lying to the blue
of the HB at $V=15.85$ and $B-V=-0.27$. This star is located at angular
distance $r=10.1$ arcmin from the projected cluster center.  This does
not exclude its cluster membership but clearly additional data are
needed to clarify this point. The equatorial coordinates of this star
are $\alpha_{2000} = 15:18:32.7$ and $\delta_{2000} = 01:54:52.8$.  The
CMD also includes   a faint blue star with $V$ = 18.836  and $B-V = -0.306$.
It is located about 2 magnitudes below the cutoff of the blue
horizontal branch of the  cluster.  This star is marked with the
triangle in Fig. 4. Its equatorial coordinates are:  $\alpha_{2000}
=15^h:18^m:14.5^s$  and $\delta_{2000} = 02^\circ:04':09.8''$.  It is a good candidate
for an extreme horizontal branch star. The light curve of this star
shows no evidence for variability with a full amplitude exceeding 0.02
mag.

\section{Conclusions}
The most interesting result of this paper is the identification of a
detached eclipsing binary which is most probably composed of two upper
main sequence stars from M5. Photometric and spectroscopic follow up
observations of this star are planned by our group.  Our data indicate
that 10 candidate eclipsing binaries reported by Reid (1996) and Yan \&
Reid (1996) do not show any periodic variability on time scales shorter
than about a week. Four SX Phe stars  were identified, all located
among the blue stragglers on the cluster CMD.  Assuming cluster
membership, their absolute magnitudes $M_{V}$  are 0.2-0.3 mag too
bright in comparison with the standard relations obtained by McNamara
(1997, 1995).  Two outbursts of the dwarf nova M5-V101 were detected
during the period May-Aug 1997.  A faint blue object, which we propose
is an extreme horizontal branch star from M5, was identified.

\begin{acknowledgements}
JK and WK were supported by the Polish Committee of Scientific
Research through grant 2P03D-011-12 and by NSF grant AST-9528096 to
Bohdan Paczy\'nski. WP was supported by the Polish Committee of
Scientific Research through grant 2P03D-010-15.
We are indebted to Dr. B. Paczy\'nski for
his long-term support to this project. We thank Dr. Samus for 
providing us with equatorial coordinates for M5 RR Lyr variables.
Dr. E.L. Sandquist kindly sent us his BVI data for M5.
Special thanks are due to Tomek Plewa for his  help with
the MINGA package. We thank Dr. Neill Reid for examining his CCD frames
in detail, and Dr. George Preston for many useful discussions.

\end{acknowledgements}

\vspace{7pt}
\clearpage
\setcounter{figure}{0}
\begin{figure}
\epsfig{file=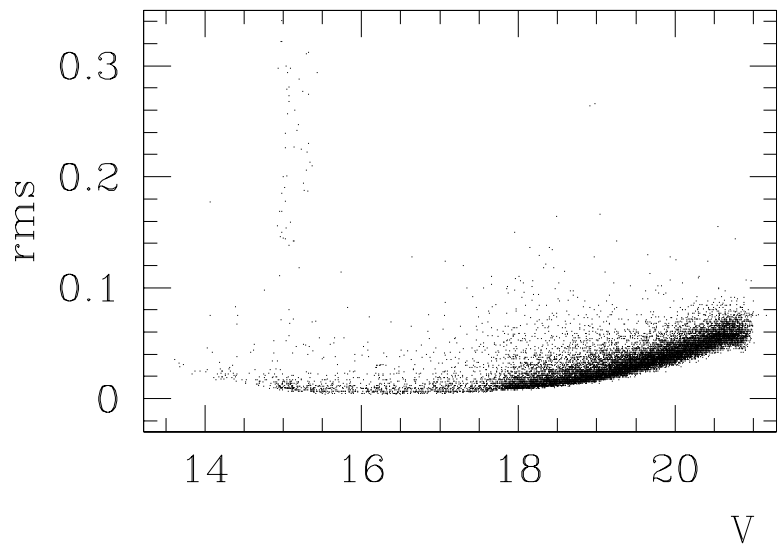, width=23cm}
\caption{Standard deviation versus average $V$ magnitude for
14847 stars from the north field. Only stars
measured on at least 75 out of the 138 best frames are marked.
}
\end{figure}
\clearpage
\setcounter{figure}{1}
\begin{figure}
 \epsfig{file=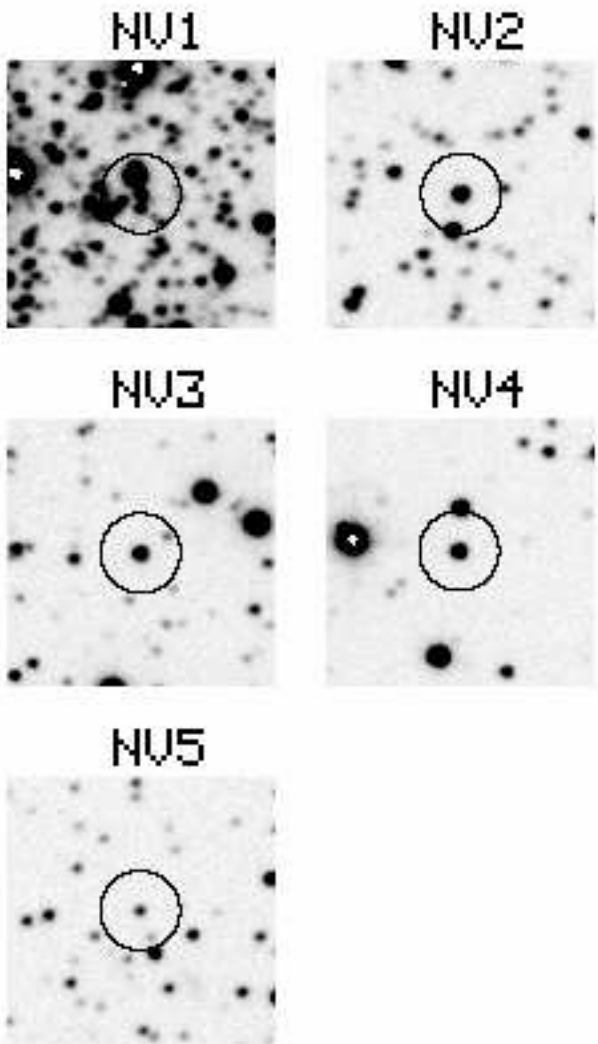}
\caption{
Finder charts for newly identified variables.
Each chart is 44 arcsec on
a side with north up and east to the left.
}
\end{figure}
\clearpage
\setcounter{figure}{2}
\begin{figure}
\epsfig{file=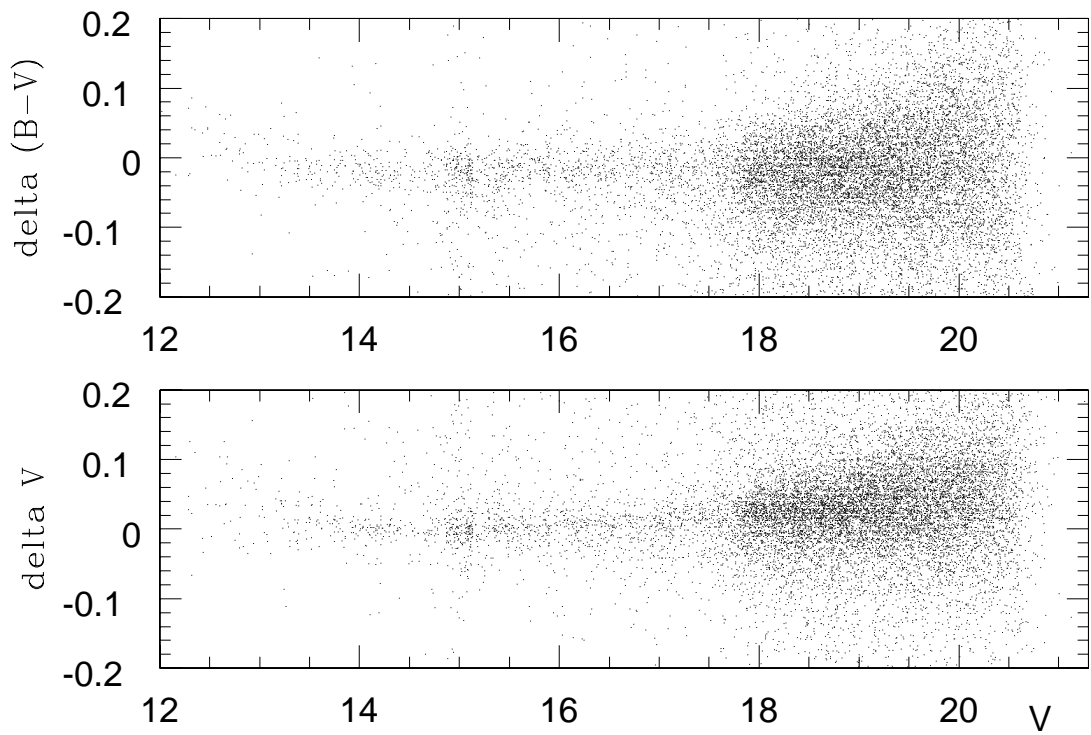}
\caption{
Comparison of our photometry with Sandquist et al. (1996).
}
\end{figure}
\clearpage
\setcounter{figure}{3}
\begin{figure}
\epsfig{file=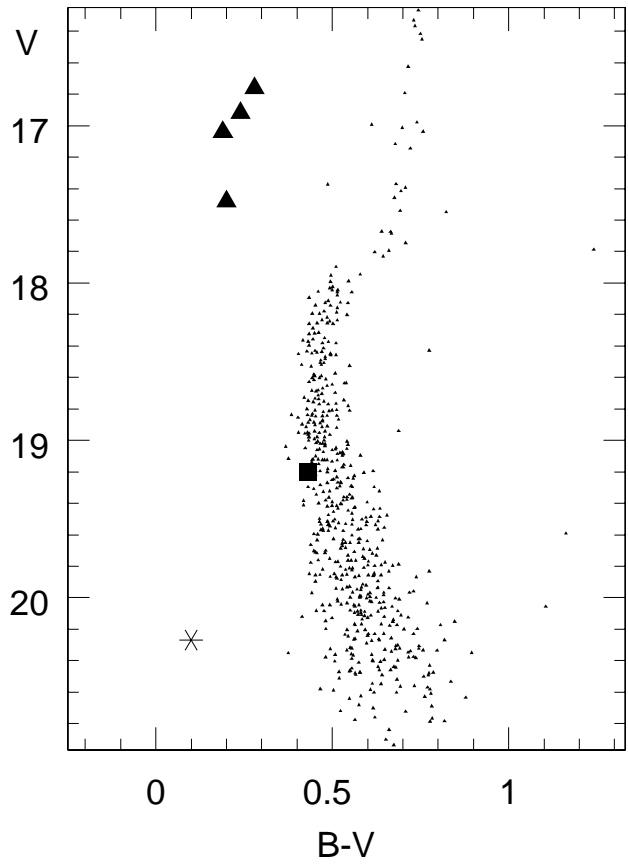}
\caption{
A $V/B-V$ CMD for M5 based on our data. Filled triangles correspond to
SX Phe stars.  The position of the eclipsing binary 
is  marked with a square.
Known RR Lyr stars are not plotted. The position of M5-V101 is marked
with an asterisk
}
\end{figure}
\clearpage
\setcounter{figure}{4}
\begin{figure}
\epsfig{file=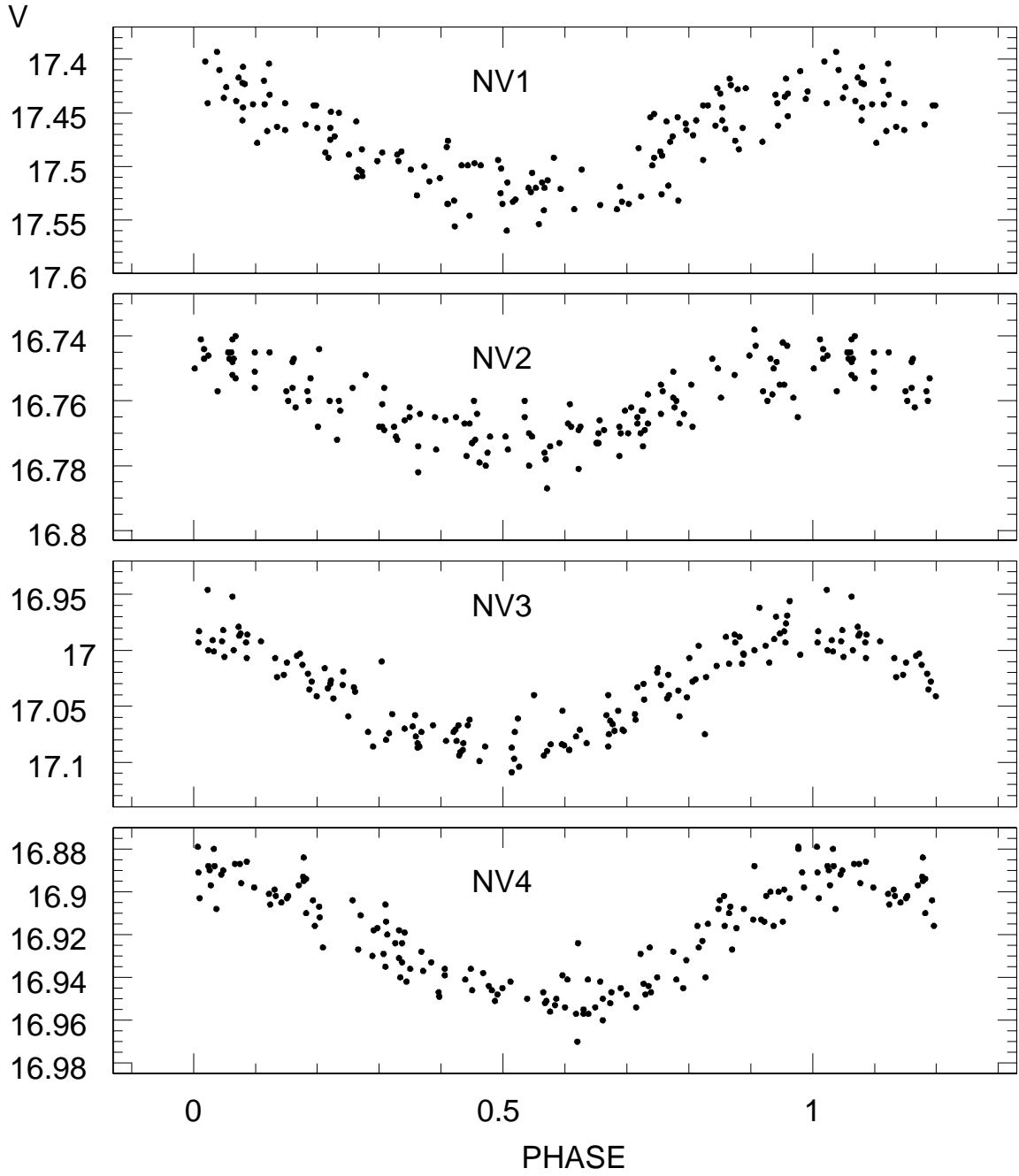}
\caption{
Phased light curves for 4 SX Phe stars from M5.
}
\end{figure}
\clearpage
\setcounter{figure}{5}
\begin{figure}
\epsfig{file=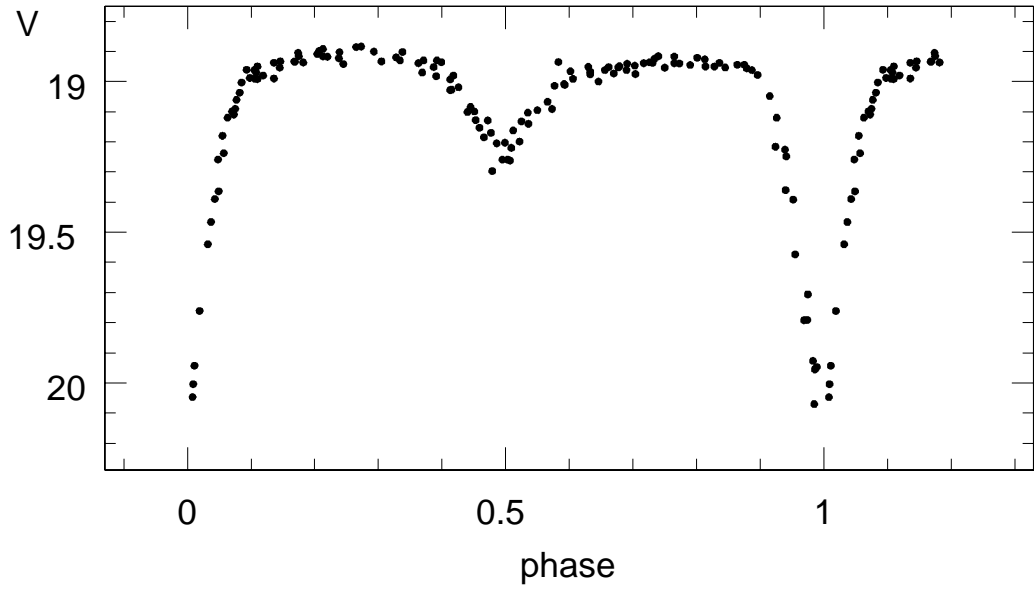}
\caption{
Phased light curve for the eclipsing binary star  M5-NV5.
}
\end{figure}
\clearpage
\setcounter{figure}{6}
\begin{figure}
\epsfig{file=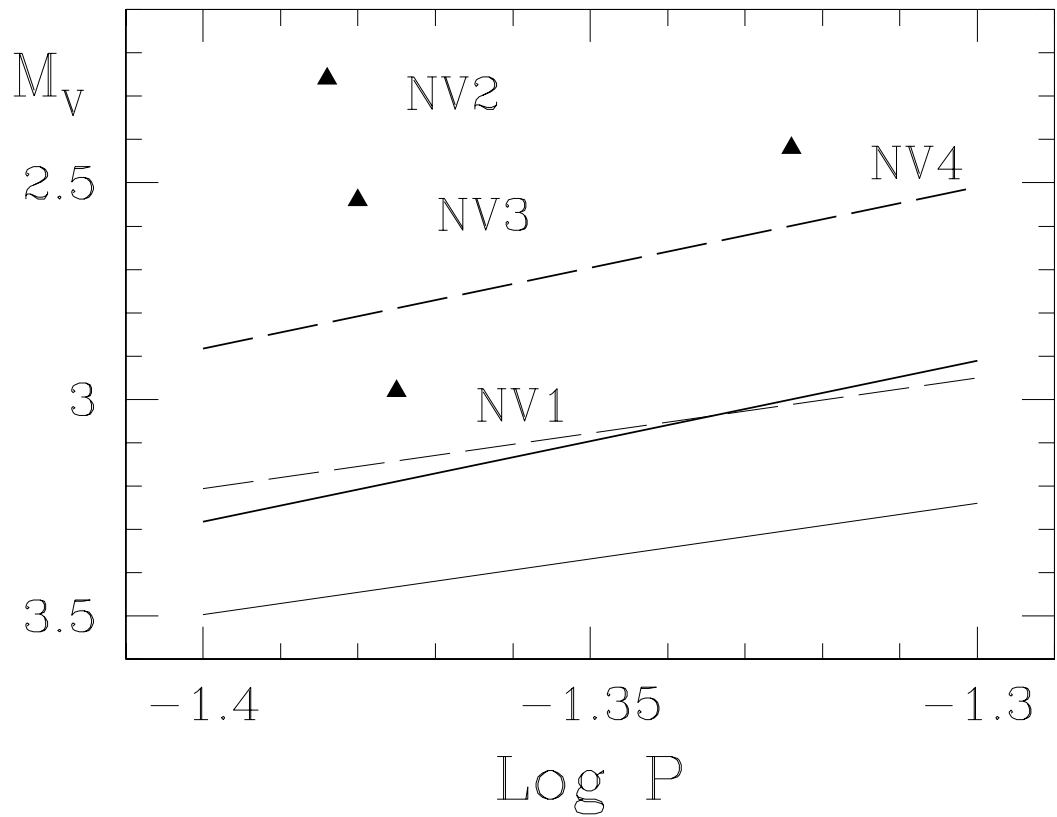}
\caption{
The period - luminosity diagram for SX Phe stars from the field of M5. Standard
relations for fundamental (solid line) and first overtone pulsators
(dashed line) are marked: M -- McNamara (1997); N -- Nemec et al. (1994).
}
\end{figure}
\clearpage
\setcounter{figure}{7}
\begin{figure}
\epsfig{file=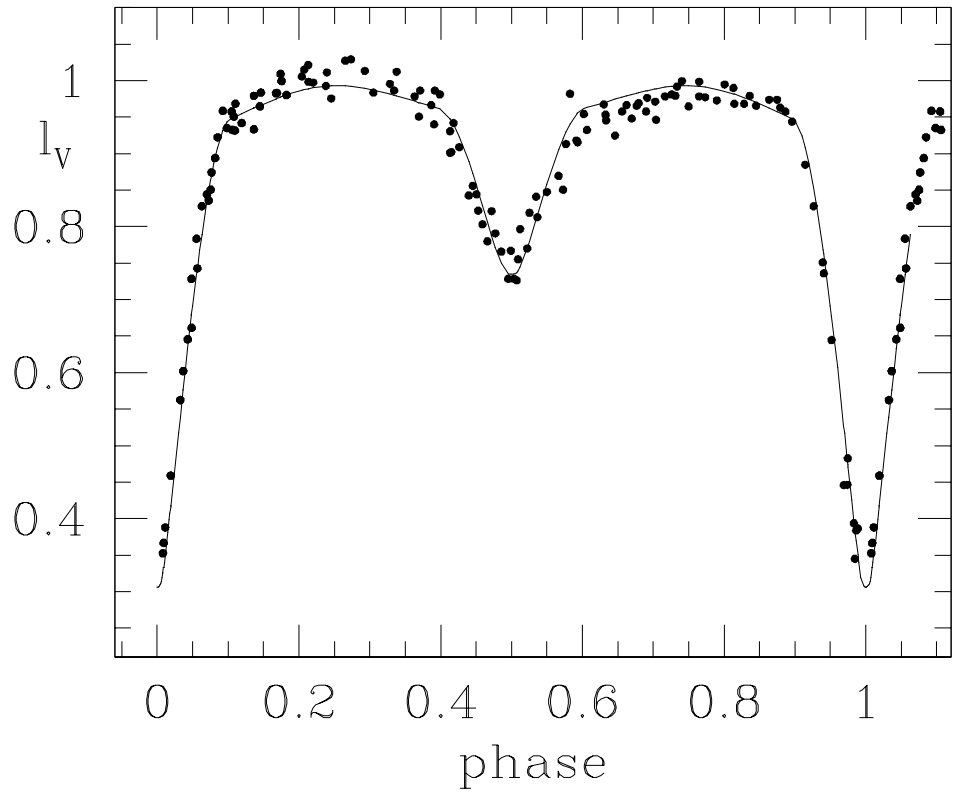}
\caption{
The observed light curve of M5-NV5  overplotted with a synthetic light curve
corresponding to an assumed mass ratio of $q=0.70$.
}
\end{figure}
\clearpage
\setcounter{figure}{8}
\begin{figure}
\epsfig{file=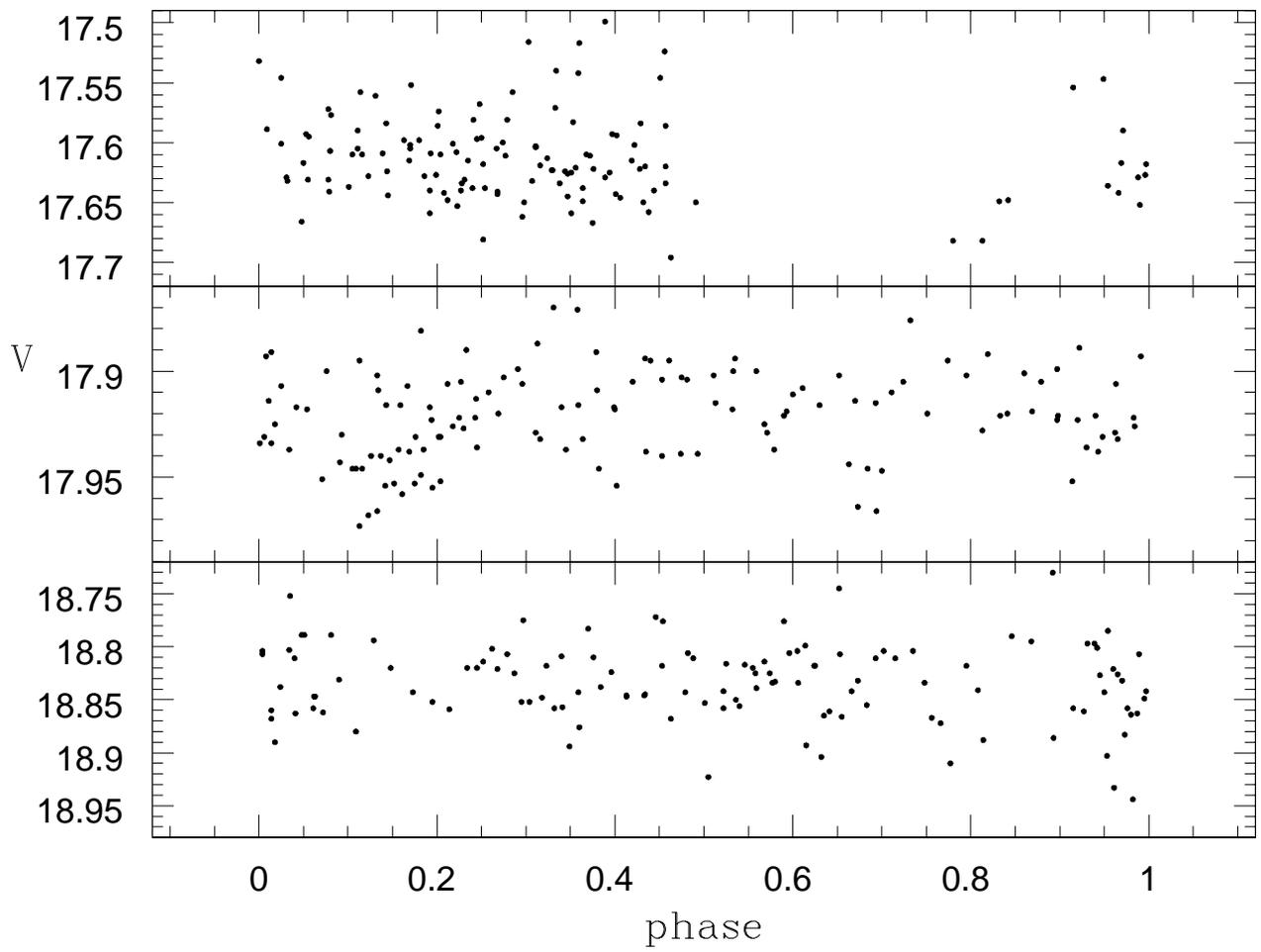}
\caption{
Observed light curves for stars V1 (top), V2 and V3 (bottom)
from Yan \& Reid (1996) phased with periods listed in that paper. 
}
\end{figure}
\clearpage
\setcounter{figure}{9}
\begin{figure}
\epsfig{file=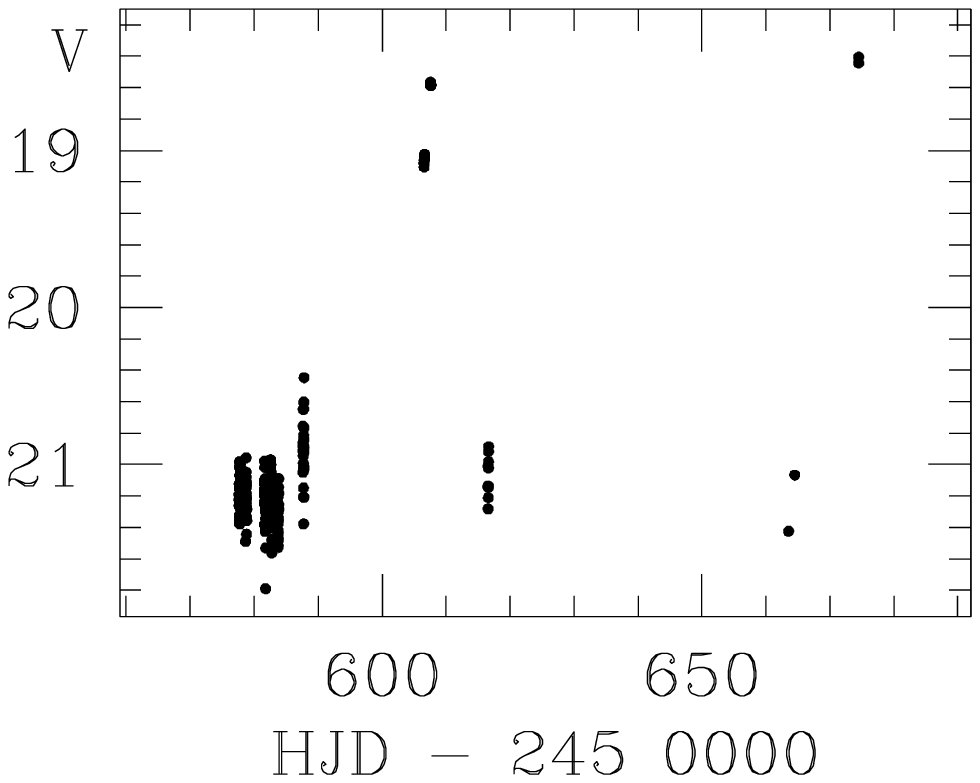}
\caption{
The $V$ band light curve of the dwarf nova M5-V101.
}
\end{figure}
\clearpage
\setcounter{figure}{10}
\begin{figure}
\epsfig{file=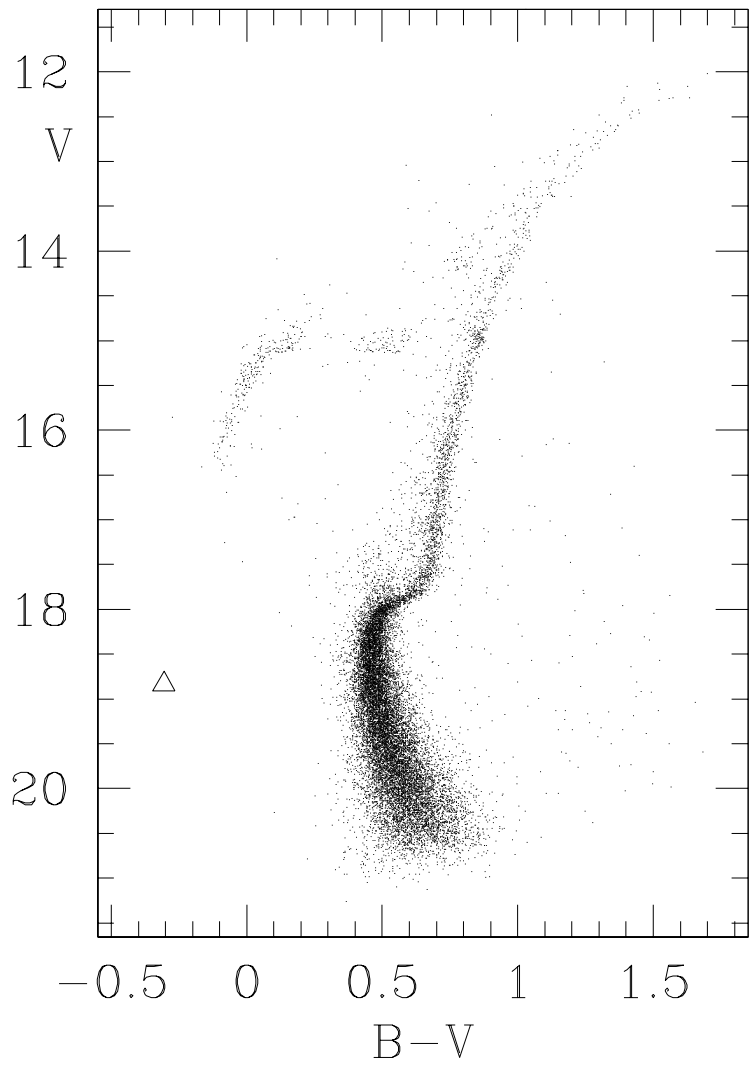}
\caption{
A V/B-V CMD for the monitored field. The trangle corresponds to 
a possible cluster extreme horizontal branch star.
}
\end{figure}
\end{document}